\documentclass[sigconf]{acmart}

\usepackage[tight,footnotesize]{subfigure}
\usepackage{url}

\usepackage{xspace} 
\usepackage{amsmath}
\usepackage{xstring} 
\usepackage{datatool,xparse}
\usepackage{algorithm}
\usepackage{algorithmic}
\usepackage{dsfont}
\usepackage{listings}
\usepackage[neverdecrease]{paralist}

\usepackage{todonotes}

\usepackage{hyperref}
\hypersetup{
	breaklinks,
    pdfpagemode={UseOutlines},
    bookmarksopen,
    pdfstartview={FitH},
    colorlinks,
    linkcolor={blue},
    citecolor={red},
    urlcolor={blue}
}

\newif\ifanonymous
\newif\ifnotanonymous
\newif\ifabridged
\newif\ifnotabridged

\notanonymoustrue

\newcommand{\setype}[1]{{\texttt{#1}\xspace}}
\newcommand{\textbfit}[1]{{\textbf{\textit{#1}}\xspace}}

\newcommand{\intel}[0]{Intel{}\xspace}
\newcommand{\paxgr}[0]{PaX/Grsecurity{}\xspace}
\newcommand{\refcountt}[0]{\setype{refcount\_t}\xspace}
\newcommand{\uintmax}[0]{\setype{UINT\_MAX}\xspace}

\newcommand{\refcountaddnotzero}[0]{\setype{refcount\_add\_not\_zero}\xspace}
\newcommand{\mpxkload}[0]{\setype{mpxk\_load\_bounds}\xspace}

\acmDOI{}

\acmISBN{}

\acmConference[]{}{}{} 
\acmYear{}
\copyrightyear{}

\acmArticle{}
\acmPrice{}

\begin{document}

\setdefaultleftmargin{0.5cm}{1cm}{}{}{}{}

\renewcommand\footnotetextcopyrightpermission[1]{}
\settopmatter{printacmref=false}
\pagestyle{plain}
\fancyfoot[C]{\thepage}

\title{Towards Linux Kernel Memory Safety}

\ifnotanonymous

\author{Elena Reshetova}
\affiliation{%
  \institution{Intel OTC Finland}
  \city{Espoo}
	\country{Finland}
}
\email{elena.reshetova@intel.com}

\author{Hans Liljestrand}
\affiliation{%
  \institution{Aalto University}
  \city{Espoo}
	\country{Finland}
}
\email{hans.liljestrand@aalto.fi}

\author{Andrew Paverd}
\affiliation{%
  \institution{Aalto University}
  \city{Espoo} 
	\country{Finland}
	}
\email{andrew.paverd@ieee.org}

\author{N.Asokan}
\affiliation{%
  \institution{Aalto University}
  \city{Espoo}
  \country{Finland}
}
\email{asokan@acm.org}

\renewcommand{\shortauthors}{E. Reshetova et al.}

\else
\author{Submission 44}
\affiliation{
}
\renewcommand{\shortauthors}{Submission 44}
\fi

%
%


\begin{abstract}
The security of billions of devices worldwide depends on the security and robustness of the mainline Linux kernel.
However, the increasing number of kernel-specific vulnerabilities, especially memory safety vulnerabilities, shows that the kernel is a popular and practically exploitable target. 
Two major causes of memory safety vulnerabilities are reference counter overflows (temporal memory errors) and lack of pointer bounds checking (spatial memory errors). 

To succeed in practice, security mechanisms for critical systems like the Linux kernel must also consider performance and deployability as critical design objectives.
We present and systematically analyze two such mechanisms for improving memory safety in the Linux kernel: (a) an overflow-resistant reference counter data structure designed to accommodate typical reference counter usage in kernel source code, and (b) runtime pointer bounds checking using Intel MPX in the kernel.

\end{abstract}

\maketitle

\section{Introduction}
The Linux kernel lies at the foundation of millions of different devices around us, ranging from servers and desktops to smartphones and embedded devices.
While there are many solutions for strengthening Linux application security covering access control frameworks (SELinux~\cite{smalley2001implementing}, AppArmor~\cite{bauer2006paranoid}), integrity protection systems (IMA/EVM~\cite{ima}, dm-verity~\cite{dm-verity}), encryption, key management and auditing, they are rendered ineffective if an attacker can gain control of the kernel.
Recent trends in Common Vulnerabilities and Exposures (CVEs) indicate a renewed interest in exploiting the kernel~\cite{nistCves}.
Kernel bugs are long-lived, on average taking five years before being found and fixed~\cite{cooklss2016}, and even when fixed, security updates might not be deployed to all vulnerable devices. 
Thus, we cannot rely solely on retroactive bug fixes, but need proactive measures to harden the kernel by limiting its exploitability.
This is the goal of the Kernel Self Protection Project~(KSPP)~\cite{kspp}, a large community of volunteers working on the mainline Linux kernel.
In this paper, we describe our contributions, as part of the KSPP, to the development of two recent kernel memory safety mechanisms.

Depending on their severity, memory errors can allow an attacker to read, write, or execute memory,
thus making them attractive targets. For example, use-after-free errors and buffer overflows feature
prominently in recent Linux kernel CVEs~\cite{raheja2016analysis, chen2011linux}.
Memory errors arise due to the lack of inherent memory safety in C, the main implementation language
of the Linux kernel. There are two fundamental classes of memory errors:

\textbf{Temporal memory errors} occur when pointers to uninitialized or freed memory are dereferenced.
One common case is a \emph{use-after-free} error e.g., dereferencing a pointer that has been prematurely freed by another execution thread.
A null pointer dereference error, although common, is more challenging to exploit~\cite{xu2015collision}, whereas use-after-free errors are both exploitable and common, and have featured in CVE-2014-2851, CVE-2016-4558, and CVE-2016-0728, among others.

\textbf{Spatial memory errors} occur when pointers are used to access memory outside the bounds of their intended areas.
A prime example is a buffer overflow, which occurs when the amount of data written exceeds the size of the buffer.
Spatial memory errors have appeared in CVE-2014-0196, CVE-2016-8440, CVE-2016-8459, and CVE-2017-7895.

While memory safety has been scrutinized for decades (Section~\ref{sec:related-work}), much of the work has focused on user-space.
These solutions are not readily transferable to kernel-space.

For instance, solutions to protect against spatial memory errors use a static and fast
addressing scheme to store pointer bounds by treating virtual memory as an endless physical memory~\cite{nagarakatte2009softbound,ramakesavan2015intel}.
This approach is not viable in kernel-space as it cannot trivially handle arbitrary page faults.
Although there have been a few proposals for improving kernel memory safety (e.g.,
kCFI~\cite{Rigo} and KENALI~\cite{kenali}), these have not considered the critical issue of
deployability in the mainline Linux kernel. Other mechanisms, such as the widely used Kernel
Address Sanitizer (KASAN)~\cite{kasan}, are intended as debugging facilities, not runtime
protection.

In this paper, we present solutions for mitigating the major causes of both temporal and spatial memory errors: 
(1) we contributed to the design of \refcountt a new reference counter data type that prevents reference counter overflows; in particular, we analyzed how developers have used reference counters in the kernel, and based on this we designed extensions to the new \refcountt API; and 
(2) we developed a mechanism for performing efficient pointer bounds checking at runtime, using Intel Memory Protection Extensions (MPX). 
The new \refcountt data structure and API are already integrated into the Linux mainline kernel. 
More than half our patches (123/233 at the time of writing) to convert existing reference counters to use \refcountt have also been already integrated.

In summary, we claim the following contributions:
\begin{itemize}
 
	\item \textbf{Extended \refcountt API}: We present a heuristic technique to identify
	 instances of reference counters in the kernel source code
         (Section~\ref{sec:analyzing-coccinelle}). By analyzing
         reference counter usage in the kernel,
	 we designed extensions to the API of the new \refcountt reference counter data
         structure (Section~\ref{sec:updating-refcount-api}). We also converted the entire kernel source tree to 
	 use the new API (Section~\ref{sec:deployment-refcount}). 

	\item \textbf{MPXK}: We present a spatial memory error prevention mechanism for the Linux kernel
	based on the recently released \intel Memory Protection Extensions (MPX)
	(Section~\ref{sec:sol-mpxk}).

	\item \textbf{Evaluation}: We present a systematic analysis of \refcountt and MPXK 
 in terms of performance, security, and usability for kernel developers. (Section~\ref{sec:evaluation}).

\end{itemize}
Our primary objective in this work was to deploy our results into the mainline Linux kernel. In Section~\ref{sec:discussion} we reflect on our experience in this regard, and offer suggestions for other researchers with the same objective.

\section{Background}
\label{sec:background}
\subsection{Linux kernel reference counters} 

Temporal memory errors are especially challenging in low-level systems that cannot rely on automated
facilities such as garbage collection for object destruction and freeing of allocated memory. In
simple non-concurrent C programs, objects typically have well defined and predictable allocation and
release patterns, which make it trivial to free them manually. However, complex systems, such as the Linux
kernel, rely heavily on object reuse and sharing patterns to minimize CPU and
memory use.

In lieu of high-level facilities like garbage collection, object lifetimes can be managed by using \textbfit{reference
counters} to keep track of current uses of an object~\cite{Collins1960}. Whenever a new reference to an
object is taken, the object's reference counter is incremented, and whenever the object is released
the counter is decremented. When the counter reaches zero,

it means that the object is
no longer used anywhere, so the object can be safely destroyed and its associated memory can be freed. However, reference counting schemes are historically error prone; A missed increment or
decrement, often in a rarely exercised code path, could imbalance the counter and cause either
a memory leak or a use-after-free error.

Reference counters in the Linux kernel are typically implemented using the \setype{atomic\_t}
type~\cite{mckenney2007overview}, which in turn is implemented as an \setype{int} with a general
purpose atomic API consisting of over 100 functions. This approach poses two problems. First,
the general purpose API provides ample room for subtly incorrect implementations of reference
counting schemes motivated by performance or implementation shortcuts.

Second, as a general purpose integer \setype{atomic\_t}

allows its instances to overflow.
When reference counters are based on \setype{atomic\_t}, this implies that they can inadvertently reach zero via only increments. Overflow
bugs are particularly hard to detect using static code analysis or fuzzing techniques because they
require many consequent iterations before the overflow happens~\cite{nikolenko2016}. A recent
surge of exploitable errors, such as CVE-2014-2851, CVE-2016-4558, CVE-2016-0728, CVE-2017-7487 and
CVE-2017-8925, specifically target reference counters.

\subsection{\intel Memory Protection Extensions (MPX)}
\label{sec:background-mpx}

\intel Memory Protection Extensions (MPX)~\cite{ramakesavan2015intel} is a recent technology to prevent spatial memory errors. 
It is supported on both Pentium core and Atom core micro-architectures from Skylake and Goldmount onwards, thus targeting a significant range of end devices from desktops to mobile and embedded processors.
In order to use the MPX hardware, both the compiler and operating system must support MPX.
On Linux, MPX is supported by the GCC and ICC compilers, and the kernel supports MPX for user-space applications.
MPX is source and binary compatible, meaning that no source code changes are required and that MPX-instrumented binaries can be linked with non-MPX binaries (e.g., system libraries). 
MPX-instrumented binaries can still be run on all systems, including legacy systems without MPX hardware.
If MPX hardware is available, the instrumented binaries will automatically detect and configure this hardware during process initialization.

MPX prevents spatial memory errors by checking pointer bounds before a pointer is dereferenced.
Conceptually, every pointer is associated with an upper and lower bound.
Pointer bounds are determined by means of compile-time code instrumentation.
For a given pointer, the bounds can either be set statically (e.g., based on static data structure sizes), or dynamically (e.g., using instrumented memory allocators).
For example, \setype{malloc} is instrumented to set the bounds of newly allocated pointers. 

In order to perform a bounds check, the pointer's bounds must be loaded in one of the four new MPX bound (\setype{bndx}) registers.
When not in these registers, the MPX instrumentation stores bounds on the stack or in static memory.
However, in some cases the number of bounds to be stored cannot be determined at compile time (e.g., a dynamically-sized array of pointers).
In these cases, bounds are stored in memory in a new two-level metadata structure, and accessed using the new bound load (\setype{bndldx}) and store (\setype{bndstx}) instructions.
These instructions use the address of the pointer to look up an entry in the process's Bound Directory (BD), which in turn points into a operating system managed Bound Table (BT), which stores the pointer's address and bounds.
On 64-bit systems the BD is 2~GB, and each individual BT is 4~MB.
To reduce this high memory overhead, the Linux kernel only allocates physical memory to the BD regions when they are written to, and only allocates individual BTs when they are accessed.

When a bounds check fails, the CPU issues an exception that must be handled by the operating system.

\section{Problem statement}
\label{sec:prob-statement}
The goal of this work is to limit the Linux kernel's vulnerability to memory safety errors, with
a focus on common errors and attacks. Specifically, we focus on preventing two causes of memory errors:

\begin{itemize}
\itemsep0em

	\item \textbf{Reference counter overflows}: the security requirement is to design
	a reference counter type and associated API such that the counter is guaranteed never to
	overflow.

	\item \textbf{Out-of-bounds memory accesses}: the security requirement is to design
	an access control scheme that prevents out-of-bounds accesses.

\end{itemize}

In addition to the security requirements, 

the following are mandatory design considerations if the solution is to be used in practice:

\begin{itemize}
\itemsep0em

	\item \textbf{Performance.} Any solutions added to the kernel must minimize performance
	impact. Some subsystems, such as networking and filesystem, are particularly sensitive to
	performance. Security mechanisms perceived as imposing a high performance
	overhead risk being rejected by the maintainers of these subsystems. Furthermore, the
	Linux kernel runs on a vast range of devices, including closed fixed-function devices like
	routers, where software attacks are not a threat but performance requirements are stringent.
	Thus, the solutions must adapt or be configurable depending on the needs of different usage
	scenarios.

	\item \textbf{Deployability.} Given our goal of integrating our solutions into the
	mainline Linux kernel, their implementation must follow the Linux
	kernel design guidelines, impose minimal kernel-wide changes, and be both maintainable and
	usable. \emph{Usability for kernel developers} is particularly crucial for new features that may be adopted at the discretion of subsystem developers.

\end{itemize}

\section{Reference Counter Overflows}
\label{sec:sol-refcounter}

We developed a methodology to
find reference counters in the kernel source code as well as analyze how they are typically used. 
A new type \refcountt and corresponding minimal API were introduced by Peter Zijlstra, one of the Linux maintainers of related subsystems.
It provides various
reference counter specific protections by preventing some outright incorrect behavior, i.e.,
incrementing on zero and overflowing the counter.
Based on our analysis of reference counter use, we proposed several additions to the initial
\refcountt API, making it widely usable as a replacement for existing reference counters and
 future implementations. To comprehensively prevent reference counter overflows, we have developed a set of patches to convert 
all conventional reference counters in the kernel to use \refcountt.

\subsection{Analyzing Linux Kernel reference counters}
\label{sec:analyzing-coccinelle}

Reference counters are spread across the Linux kernel. We thus systematically analyzed the source
 code to locate instances of reference counters based on the \setype{atomic\_t} type using Coccinelle~\cite{coccinelle}, a static
analyzer integrated into the Linux Kernel build system (KBuild). Coccinelle takes code patterns as
input and finds (or replaces) their occurrences in the given code base. We defined three
such code patterns to identify reference counters based on their behavior (see Listing~\ref{cocci} in the Appendix for the full patterns):

\begin{enumerate}

	\item Freeing a referenced object based on the return value from \setype{atomic\_dec\_and\_test}
	or one its variants. This is the archetypical reference counter use case. 

	\item Using \setype{atomic\_add\_return} to decrement a variable and compare its updated value,
	typically against zero. These use cases are essentially variations of the basic
	\setype{dec\_and\_test} cases albeit using a different function for the implementation.
	
	\item Using \setype{atomic\_add\_unless} to decrement a counter only when its
	value differs from one. This case is less common.

\end{enumerate}

These patterns are strong indicators that the identified object employs an \setype{atomic\_t} reference
counting scheme. So far, this approach detected all occurrences of reference counters. Some
false positives were reported, particularly on implementations with \setype{atomic\_t} variables that 
occasionally exhibit reference counter-like behavior.
For example, under one condition an object might be freed when such counter reaches zero (behaves like a reference counter),
but under a different condition the object might instead be recycled when the counter reaches zero (see Section~\ref{sec:sol-refcounter-chal} for examples).

Of the 250 \setype{atomic\_t} variables reported on an
unmodified v4.10 kernel we have manually confirmed 233 variables as reference counters.

\subsection{refcount\_t API}
\label{sec:refcount-api}

The initial \refcountt API, introduced by Peter Zijlstra%
\ifnotanonymous
\footnote{\url{http://lwn.net/Articles/713645/}}
\else
\footnote{Anonymized}
\fi
(Listing~\ref{initial}), was designed around strict semantically correct reference
counter use. This meant that beyond \setype{set} and \setype{read} calls, the API provided only two
incrementing functions, both of which refuse to increment the counter when its value is zero, and three variations of
decrementing functions that enforce the checking of return values via compile time warnings. When
decrementing the caller needs to know if the value reached zero, which is indicated by a return
value of \setype{true}. A \setype{true} return value from \setype{refcount\_inc\_not\_zero} means that the counter was
non-zero, indicating that the referenced object is safe to use.

\lstset{language=C, basicstyle=\ttfamily\footnotesize, 
	 basicstyle=\footnotesize\ttfamily,
	 commentstyle=\itshape\color{purple!40!black},
	 identifierstyle=\color{blue},
	 escapeinside={(*@}{@*)},
	 stringstyle=\color{orange}
	}

\lstset{language=C++, basicstyle=\ttfamily\footnotesize}
\begin{lstlisting}[label=initial,caption=Initial bare \refcountt API.]
void refcount_set(refcount_t *r, unsigned int n);
unsigned int refcount_read(const refcount_t *r);
bool refcount_inc_not_zero(refcount_t *r);
void refcount_inc(refcount_t *r);
bool refcount_dec_and_test(refcount_t *r);
bool refcount_dec_and_mutex_lock(refcount_t *r,
                                 struct mutex *lock);
bool refcount_dec_and_lock(refcount_t *r,
                           spinlock_t *lock);
\end{lstlisting}

\begin{figure}[t]
\centering
\includegraphics[width=1.7in]{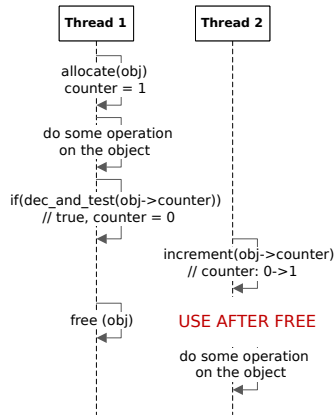}
	\caption{Potential use-after-free when incrementing a reference counter from zero.}
\label{fig:inc-on-zero}
\end{figure}

The
checks are performed to ensure that a reference counter reaching zero triggers object release, thus
preventing a potential memory leak, whereas increment on zero is prohibited to avoid potential
use-after-free, as illustrated in Figure~\ref{fig:inc-on-zero}.

The main challenge in designing the \refcountt API was how to deal with an event that would otherwise
cause the counter to overflow. One approach would be to simply ignore the event, such that the
counter remains at its maximum value. However, this means that the number of references to the
object will be greater than the maximum counter value. If the counter is subsequently decremented,
it will reach zero and the object will be freed before all references have been released (i.e.,
leading to a use-after-free error). To overcome  this challenge, the \refcountt API instead
\emph{saturates} the counter, such that it remains at its maximum value, even if there are
subsequent increment or decrement events. A saturated counter would, therefore, result in a memory
leak since the object will never be freed. However, a cleanly logged memory leak is a small price to
pay for avoiding the potential security vulnerabilities of a reference counter overflow. This
approach is similar to that previously used by the \paxgr patches~\cite{grsecOnRefcount}. 

\subsection{Our extensions to the \refcountt API}
\label{sec:updating-refcount-api}

Our efforts in finding existing reference counters revealed several
variations of the strict archetypical reference counting schemes, which are incompatible with
the initial \refcountt API\@. As a result, we designed several API additions that got integrated into the Linux kernel%
\ifnotanonymous
\footnote{\url{http://www.openwall.com/lists/kernel-hardening/2016/11/28/4}}
\else
\footnote{Anonymized}
\fi. 
The new API calls are
shown in Listing~\ref{addition}. 

\lstset{language=C++, basicstyle=\ttfamily\footnotesize}
\begin{lstlisting}[label=addition, caption=Our additions to the \refcountt API.]
void refcount_add(unsigned int i, refcount_t *r);
bool refcount_add_not_zero(unsigned int i, refcount_t *r);
void refcount_dec(refcount_t *r);
bool refcount_dec_if_one(refcount_t *r);
bool refcount_dec_not_one(refcount_t *r);
void refcount_sub(refcount_t *r);
bool refcount_sub_and_test(unsigned int i, refcount_t *r);
\end{lstlisting}

The functions allowing arbitrary additions and subtractions, i.e., the \setype{refcount\_add} and
\setype{refcount\_sub} variants, are needed in situations where larger value changes are needed.
For example, the \setype{sk\_wmem\_alloc} variable in the networking subsystem serves as a reference
counter but also tracks transfer queue, which need arbitrary additions and
subtractions.  
\setype{refcount\_sub\_and\_test} and \setype{refcount\_add\_not\_zero} also provide a return value that,
that indicate that the counter reached the value of zero.
\setype{refcount\_dec} accommodates situations where the forced return value checks would incur
needless overhead.
For instance, some functions in the \setype{btrfs} filesystem handle nodes that are guaranteed to be
cached, i.e., at least the reference held by the cache will be remaining.
Finally, \setype{refcount\_dec\_if\_one} and
\setype{refcount\_dec\_not\_one} enable schemes that require specific operations before or
instead of releasing objects. For instance, the networking subsystem extensively utilizes patterns
where the value of one indicates that an objects is invalid, but can be recycled.

\subsection{Implementation considerations}
\label{sec:refcount-implementation}

To avoid costly lock or mutex use the \refcountt API generic, i.e., not architecture specific,
implementation uses the
\emph{compare-and-swap} pattern, which is built around the atomic \setype{atomic\_cmpxchg}
function, shown in Algorithm~\ref{code:cmpxchg}. On x86 \setype{atomic\_cmpxchg} is implemented as a
a single atomic CPU instruction, but the implementation is guaranteed to be atomic regardless
of architecture. The function always returns the prior value but exchanges it only if it was equal
to the given condition value $\mathit{comp}$. The compare-and-swap works by indefinitely looping until a
\setype{cmpxchg} succeeds. This avoids costly locks and allows all but the \setype{cmpxchg} to be
non-atomic. Note that in the typical case, without concurrent modifications, the loop runs only
once, thus being much more efficient than locks and potential blocking of execution.

\begin{algorithm}\caption{\setype{cmpxchg}($\mathit{atomic}$, $\mathit{comp}$, $\mathit{new}$) sets the value of $atomic$ to
$new$ if, and only if, the prior value of $\mathit{atomic}$ was equal to $\mathit{comp}$. Whether the value was
changed or not the function always returns the prior value of $\mathit{atomic}$. In practice the
function is implemented as a single inline CPU instruction.}
	\label{code:cmpxchg}
	\begin{algorithmic}[1]
		\STATE $old \leftarrow atomic.value$
		\IF{$old = comp$}
			\STATE $atomic.value \leftarrow new$
		\ENDIF
		\RETURN $old$
	\end{algorithmic}
\end{algorithm}

As an example of the extended \refcountt API implementation, consider the \refcountaddnotzero function shown
in Algorithm~\ref{code:refcount-inc}. It is used to increase \refcountt when acquiring
a reference to the associated object, a return value of \setype{true} indicates that the counter was
non-zero, and thus the associated object is safe to use. The actual value from a user perspective is
thus irrelevant and \refcountaddnotzero guarantees only that the return value is true if, and only
if, the value of \refcountt at the time of the call was non-zero. Internally the function further
guarantees that the increment takes place only when the prior value was in the open interval $(0,
UINT\_MAX)$, thus preventing use-after-free due to increment from zero and use-after-free due to
overflow. This case results in the default return statement at
line~\ref{code:refcount-add-return-default}. Attempted increment from zero or \uintmax results in
returns at lines~\ref{code:refcount-add-if-zero} and~\ref{code:refcount-add-if-saturated},
respectively. Finally, an addition that would overflow the counter instead saturates it by setting
its value to \uintmax on line~\ref{code:refcount-add-do-saturate}.

\begin{algorithm}\caption{\refcountaddnotzero($\mathit{refcount}$, $\mathit{summand}$) attempts to
add a $\mathit{summand}$ to the value of $\mathit{refcount}$, and returns true if, and only if, the prior value was non-zero.
Note that the function is not locked and only the \setype{cmpxchg}
(line~\ref{code:refcount-inc-cmpxchg}) is atomic, i.e.,  the value of $\mathit{refcount.value}$
can change at any point of execution. The loop ensures that the \setype{cmpxchg} eventually
succeeds. Overflow protection is provided by checking if the value is already saturated
(line~\ref{code:refcount-add-if-saturated}) and by saturating, instead of overflowing, on
line~\ref{code:refcount-add-do-saturate}.}

	\label{code:refcount-inc}
	\begin{algorithmic}[1]
		\ENSURE{$retval = \TRUE \Leftrightarrow refcount.value > 0$}
		\ENSURE{$\mathit{value\_unchanged} \Leftrightarrow refcount.value \in \{0, \mathit{UINT\_MAX}\}$}
		\STATE $\mathit{val} \leftarrow \mathit{refcount.value}$ \COMMENT{use local copy}
		\WHILE{\TRUE}
			\IF{$\mathit{val} = 0$}
				\label{code:refcount-add-if-zero}
				\RETURN{\FALSE} \COMMENT{counter not incremented from zero}
				\label{code:refcount-add-if-zero-after}
			\ENDIF
			\IF{$\mathit{val} = \mathit{UINT\_MAX}$}
				\label{code:refcount-add-if-saturated}
				\RETURN{\TRUE} \COMMENT{counter is saturated, thus not zero}
				\label{code:refcount-add-if-saturated-after}
			\ENDIF
			\STATE $\mathit{new} \leftarrow \mathit{val} + \mathit{summand}$ \COMMENT{calculate new value}
			\label{code:refcount-inc-check-overflow}
			\IF{$\mathit{new} < \mathit{val}$}
				\label{code:refcount-add-do-saturate}
				\STATE $\mathit{new} \leftarrow \mathit{UINT\_MAX}$ \COMMENT{Saturate instead of overflow}
				\label{code:refcount-add-do-saturate-after}
			\ENDIF
			\STATE $\mathit{old} \leftarrow \mathit{atomic\_cmpxchg}(\mathit{refcount.atomic}, \mathit{val}, \mathit{new}$)
			\label{code:refcount-inc-cmpxchg} 
			\IF[if $\mathit{refcount.value}$ was unchanged, then]{$old = val$}
				\STATE \textbf{break} \COMMENT{value was updated by cmpxchg}
			\ENDIF
			\STATE $\mathit{val} \leftarrow \mathit{old}$ \COMMENT{update $\mathit{val}$ for next iteration}
		\ENDWHILE

		\RETURN{\TRUE} \COMMENT{value incremented}
		\label{code:refcount-add-return-default}
	\end{algorithmic}
\end{algorithm}

\subsection{Challenges}
\label{sec:sol-refcounter-chal}

Despite the additions to the \refcountt API, several reference counting instances 
nonetheless required careful analysis and in some cases challenging modifications to the underlying
logic. These challenges were the main reason for not using Coccinelle to automatically convert reference counting schemes.
The most common challenges are \emph{object pool patterns} and unconventional reference counters.

While the object pool pattern is accommodated by our additions to the \refcountt API
- namely, by providing functions that distinguish the value one - such implementations typically
employ negative values or other means to distinguish between freed and recyclable objects. These
patterns therefore often necessitated non-trivial changes to ensure that neither increment on zero
operations nor negative values are expected. Overall we encountered 6 particularly challenging
recycling schemes.  For example, the \setype{inode} \refcountt conversion spanned a total of 10
patches%
\ifnotanonymous
\footnote{\url{http://lkml.org/lkml/2017/2/24/599}}
\else
\footnote{Anonymized}
\fi 
and required so many changes that it has not yet been accepted by the maintainer.

In some cases reference counters were used in \textit{non-conventional ways}, such as to govern other behavior or
track other statistics in addition to the strict reference count itself (e.g., the network socket \setype{sk\_wmem\_alloc} variable mentioned above).

The \refcountt API can be
surprising or outright erroneous for such unconventional uses; it might for instance be expected
that such variables can be incremented from zero or potentially reach negative values. Our
conversion efforts touched upon 21 reference counters in this category.

\subsection{Deployment of refcount\_t}
\label{sec:deployment-refcount}

We developed 233 distinct kernel patches, each
converting one distinct variable, spanning all the kernel subsystems. During our work, the
\refcountt API, with our additions, was also finalized in the mainline Linux kernel. The next stage of our work consisted
of submitting all the patches to the respective kernel maintainers and adapting them based on their
feedback. Based discussions with maintainers, some patches were permanently
dropped, either because they would require extensive changes in affected subsystems or would incur
unacceptable performance penalty without any realistic risk of actually overflowing the particular
counter. As explained in Section~\ref{sec:prob-statement}, some performance-sensitive systems
proved challenging to convert due to performance concerns. As a result, a new
\setype{CONFIG\_REFCOUNT\_FULL} kernel configuration option was added to allow switching the
\refcountt protections off and thus use the new API without any performance overhead. This can be
utilized by devices that have high performance requirements but are less concerned about security based
on their nature (e.g., closed devices such as routers that do not allow installation of untrusted
software).  These patching efforts, discussions, and patch reviews also uncovered related reference
counting bugs that were fixed in subsequent kernel
patches\footnote{\url{http://lkml.org/lkml/2017/6/27/567}, \url{http://lkml.org/lkml/2017/3/28/383}
etc.}.

\section{Out-of-Bounds Memory Access}
\label{sec:sol-mpxk}

To prevent spatial memory errors in the Linux kernel, we have adapted \intel MPX for in-kernel use.
MPX was designed for both user-space and kernel-space (e.g., the hardware includes separate configuration registers for each).
However, until now, the Linux kernel and the GCC compiler only supported MPX for user-space applications.

Our solution, \emph{MPX for Kernel (MPXK)}, is realized as a new GCC instrumentation plugin (based on the existing MPX support in GCC).
In the following subsections, we describe the various challenges of using MPX in the Linux kernel, and our respective solutions in MPXK.

\subsection{Memory use}
\label{sec:sol-mpxk-bounds-storage}

The first challenge arises from the high memory overhead incurred by the MPX Bound Directory (BD) and Bound Tables (BT).
As explained in Section~\ref{sec:background-mpx}, user-space MPX attempts to reduce this overhead by allocating this memory only when needed.
However, this requires the kernel to step in at arbitrary points during execution to handle page faults or Bound Faults caused by caused by BD dereferences or unallocated BTs.
This approach cannot be used in MPXK because the kernel cannot handle page faults at arbitrary points within its own execution.
It is also not feasible to pre-allocate the BD and BTs, as this would increase base memory usage by over $500\%$ and require extensive modifications to accommodate certain classes of pointers (e.g., pointers originating from user-space).
An alternative approach would be to substitute the hardware-backed BD and BTs with our own metadata, similar to SoftBound~\cite{nagarakatte2009softbound} or KASAN~\cite{kasan}.
However, this would still incur the same memory and performance overheads as those systems.

To overcome this challenge, MPXK dynamically determines pointer bounds using \emph{existing kernel metadata}.  
Specifically, we re-use the kernel memory management metadata created by \setype{kmalloc}-based allocators.
We define a new function, \mpxkload, that uses this existing metadata to determine the bounds for a pointer allocated by \setype{kmalloc}, and loads these into the MPX registers.
A side-effect of this approach is that bounds are rounded up to the nearest allocator cache size (i.e., may be slightly larger than the requested allocation size).
However, this has no security implications because the allocator will not allocate any other objects
in the round-up memory area~\cite{akritidis2009baggy}. MPXK thus never uses the \setype{bndldx} or
\setype{bndstx} instructions.
Our kernel instrumentation only uses \mpxkload when the bounds cannot be determined at compile-time
(e.g., for dynamically allocated objects).

\subsection{Kernel support code}

In user-space, MPX is supported by GCC library implementations for various support functionality, such as initialization and function wrappers. 
The user-space instrumentation initializes MPX during process startup by allocating the BD in
virtual memory and initializing the MPX hardware.
However, this existing initialization code cannot be used directly in the Linux kernel because
kernel-space MPX must be configured during the kernel boot process.

In user-space, the compiler also provides instrumented wrapper functions for all memory manipulation functions, such as \setype{memmove}, \setype{strcpy}, and \setype{malloc}. 
These user-space wrappers check incoming pointer bounds and ensure that the correct bounds are associated with the returned pointers.
They are also responsible for updating the BD and BTs (e.g., \setype{memcopy} must duplicate any BD and BT entries associated with the copied memory).
However, these user-space wrappers also cannot be used in the kernel because the kernel implements its own standard library. 

MPXK includes new code to initialize the MPX hardware during the kernel boot process. 
This is done by writing the MPX configuration into each CPU's \setype{bndcfgs} Machine Specific Register (MSR).
Although MPXK does not use a BD, it still reserves an address space for the BD but does not back this up with physical memory.
This is done to ensure that any erroneous and/or malicious invocations of the \setype{bndldx} or
\setype{bndstx} instructions will cause page faults and crash the kernel instead of potentially
overwriting arbitrary memory.

MPXK includes a new set of kernel-specific wrapper functions, which are implemented as normal in-kernel library functions, and used by our new MPXK GCC-plugin.
The MPXK wrappers are substantially less complex (and thus easier to audit for security) than their user-space counterparts, because the MPXK wrappers do not need to include logic for updating the DB or BTs.

\subsection{Binary compatibility (Mixed code)}

MPXK, like MPX, is binary compatible and can therefore be used in mixed environments with both MPXK enabled code and legacy (non-instrumented) code. 
A fundamental problem for any binary-compatible bounds checking scheme is that the instrumentation \emph{cannot track pointer manipulation performed by legacy code} and therefore cannot make any assumptions about the pointer bounds after the execution flow returns from the legacy code. 
MPX offers a partial solution by storing the pointer's value together with its bounds (using the \setype{bndstx} instruction) before entering legacy code.
When the legacy code returns and \setype{bndldx} is used to load the bounds again, this instruction will reset the pointer bounds, i.e. making them essentially infinite, if it detects that the current pointer value has been changed.
However, MPXK does not use the \setype{bndstx} and \setype{bndldx} instructions but instead always attempt to load such bounds using \mpxkload. Note that this does give an additional advantage compare to MPX, since MPXK is able to load the bounds based on the pointer value.  

Function arguments, both in MPX and MPXK, rely on the caller to supply bounds. 
As such, these schemes cannot determine bounds for arguments when called from non-instrumented code. 
In both MPX and MPXK, these bounds therefore cannot be checked.
As future work, we are investigating how to determine such bounds using the new MPXK bound load function described above.

\subsection{Kernel instrumentation}

The MPXK instrumentation is based on the existing MPX support in GCC, but uses a new GCC plugin to adapt this for use in the kernel. 
As described above, this new plugin instruments the kernel code with the new MPXK bound loading function, MPXK initialization code, and kernel-space wrapper functions. 
This plugin uses the GCC plugin system that has been incorporated into Kbuild, the Linux build system, since Linux v4.8.
This means that MPXK is seamlessly integrated with the regular kernel build workflow.
A developer simply needs to add predefined MPXK flags to any Makefile entries in order to include MPXK instrumentation.
The plugin itself is implemented in four compiler passes, of which the first three operate on the high-level intermediate representation, GIMPLE, and the last on the lower-level RTL, as follows:

\begin{itemize}

	\item \textit{mpxk\_pass\_wrappers} replaces specific memory-altering function calls with their
		corresponding MPXK wrapper functions, e.g., replacing \setype{kmalloc} calls
		\setype{mpxk\_wrapper\_kmalloc} calls.

	\item \textit{mpxk\_pass\_cfun\_args} inserts MPXK bound loads for function arguments where bounds are
		not passed via the four \setype{bndx} registers. This naturally happens when more than four
		bounds are passed, or due to implementation specifics for any argument beyond the
		sixth.

	\item \textit{mpxk\_pass\_bnd\_store} replaces \setype{bndldx} calls with MPXK bound loads, and
		removes \setype{bndstx} calls. This covers all high-level (GIMPLE) loads and saves, including
		return values to legacy function calls.

	\item \textit{mpxk\_pass\_sweeper} is a final low-level pass that removes any remaining
		\setype{bndldx} and \setype{bndstx} instructions. This pass is required to remove
		instructions that are inserted during the expansion from GIMPLE to RTL.

\end{itemize}

\section{Evaluation}
\label{sec:evaluation}
We evaluate our proposed solutions against the requirements defined in Section~\ref{sec:prob-statement}.

\subsection{Security guarantees}

We analyse the security guarantees of both \refcountt and MPXK, first through a principled theoretical analysis, and second by considering the mitigation of real-world vulnerabilities.

\subsubsection*{Reference counter overflows}

With the exception of \setype{refcount\_set} and \setype{refcount\_read}, all functions that modify
\refcountt can be grouped into \emph{increasing} and \emph{decreasing} functions. All increasing
functions maintain the invariants that i) \textbf{the resulting value will not be smaller than the
original} and that ii) \textbf{a value of zero will not be increased}. The decreasing functions
maintain corresponding invariants that i) \textbf{the resulting value will not be larger than the
original} and that ii) \textbf{a value of $UINT\_MAX$ will not be decreased}.

For example, the increasing function \setype{refcount\_add\_not\_zero} (Algorithm~\ref{code:refcount-inc}) maintains the invariants as follows:
\begin{itemize}
\item \textbf{input} $= 0$: Lines~\ref{code:refcount-add-if-zero}-~\ref{code:refcount-add-if-zero-after} prevent the counter being increased.

\item \textbf{input} $= UINT\_MAX$: Lines~\ref{code:refcount-add-if-saturated}-~\ref{code:refcount-add-if-saturated-after} ensure the counter will never overflow.

\item \textbf{input} $\in (0,UINT\_MAX)$: Lines~\ref{code:refcount-add-do-saturate}-~\ref{code:refcount-add-do-saturate-after} ensure that the counter value cannot overflow as a result of addition.
\end{itemize}

Line~\ref{code:refcount-inc-cmpxchg} ensures that the addition is performed atomically, thus preventing unintended effects if interleaved threads update the counter concurrently.
Regardless of how the algorithm exits, the invariant is maintained.
The same exhaustive case-by-case enumeration can be used to prove that all other \refcountt functions maintain the invariants.

An attacker could still attempt to cause a use-after-free error by finding and invoking an extra decrement (i.e., decrement without a corresponding increment).
This is a fundamental issue inherent in all reference counting schemes.
However, the errors caused by the extra decrement would almost certainly be detected early in development or testing.
In contrast, missing decrements are very hard to detect through testing as they may require millions of increments to a single counter before resulting in observable errors.
Thanks to the new \refcountt, missing decrements can no longer cause reference counter overflows.

In terms of real-world impact, \refcountt would have prevented several past exploits, including CVE-2014-2851, CVE-2016-0728, and CVE-2016-4558.
Although it is hard to quantify the current (and future) security impact this will have on the kernel, observations during our conversion efforts support the intuition that the strict \refcountt API discourages unsafe implementations.
For example, at least two new reference counting bugs\footnote{\url{http://lkml.org/lkml/2017/6/27/409}, \url{http://lkml.org/lkml/2017/3/28/383}} were noticed and fixed due to their incompatibility with the new API\@.

\subsubsection*{Out-of-bounds memory access}

The objective of MPXK is to prevent spatial memory errors by performing pointer bounds checking.
Specifically, for objects with known bounds, MPXK will ensure that pointers to those objects cannot be dereferenced outside the object's bounds (e.g., as would be the case in a classic buffer overflow).
A fundamental limitation of bounds checking schemes is that there are various cases in which the correct object bounds cannot be (feasibly) known by the scheme.
We enumerate each of these cases and show that MPXK is at least as secure as existing schemes.

\textbf{Pointer manipulation:}
If the attacker can corrupt a pointer's value to point to a different object \emph{without dereferencing} the pointer, this can be used to subvert bounds checking schemes.
For example, object-centric schemes such as KASAN enforce bounds based on the pointer's value.
If this value is changed to point within another object's bounds, the checks will be made (incorrectly) against the latter object's bounds.
In theory, pointer-centric schemes such as user-space MPX should not be vulnerable to this type of attack, since they do not derive bounds based on the pointer's value.
However, presumably for compatibility reasons, if MPX detects that a pointer's value has changed, it \emph{resets} the pointer's bounds (i.e., allows it to access the full memory space). 
MPXK, like KASAN, will use the corrupted value to infer an incorrect set of bounds.
This is therefore no worse than MPX or object-centric schemes.
However, it must be noted that this type of attack requires the attacker to have a \emph{prior exploit} to corrupt the pointer in the first place.

\textbf{Legacy code:}
Binary-compatibility is a fundamental problem for any scheme that tracks pointers' bounds.
This is manifest in two cases: i) pointers returned from legacy (non-instrumented) code to an instrumented caller, and ii) pointers passed from legacy code as arguments to an instrumented function.
In both cases, the pointer bounds are not known and thus cannot be tracked by the instrumented code.
These issues often cannot be addressed in binary-compatible systems, MPXK however, in case i), uses its \mpxkload function to determine and load the bounds.
One potential solution for case ii) is to add meta-data to track the bounds from legacy callers.
This could be done by relaxing the strict binary-compatibility requirement, which may be feasible for the kernel, since the whole code base is typically compiled at once.
The compiler could perform kernel-wide analysis while still only adding instrumentation to the intended subsystems.
We plan to investigate this as future work.

One current limitation of MPXK is that the \mpxkload function currently can only retrieve the bounds of objects allocated by \setype{kmalloc}.
As future work, we plan to extend this to include pointers into static memory or the stack, which are not associated with any allocation pool.
Even if MPXK is unable to determine the precise bounds, these pointers could still be restricted to sensible memory areas (e.g., specific stack frames).

As a practical demonstration of MPXK real-world effectiveness, we have tested MPXK against an exploit built around the recent CVE-2017-7184.
The vulnerability, which affects the IP packet transformation framework \setype{xfrm}, is a classic buffer overflow caused by omission of an input size verification.
We first confirmed that we can successfully gain root privileges on a current Ubuntu $16.10$ installation running a custom-built v4.8 kernel using the default Ubuntu kernel configuration.
We then recompiled the kernel applying MPXK on the \setype{xfrm} subsystem, which caused the exploit to fail with a bound violation reported by MPXK\@.

\subsection{Performance}
\label{sec:ev-perf}

\subsubsection*{Reference counter overflows}

Although the \refcountt functions consist mainly of low-overhead operations (e.g., additions and subtractions), they are often used in performance-sensitive contexts. 
Therefore, while absolute overhead of individual calls can be illuminating, we need to estimate the practical impact on overall performance.
We performed various micro-benchmarks of the individual functions during the \refcountt development\footnote{\url{http://lwn.net/Articles/718280/}}.
As shown in Table~\ref{tab:refcount-t-cpu}, \setype{refcount\_inc} introduces an average overhead of 20 CPU cycles, compared to \setype{atomic\_inc}.
However micro-benchmarks cannot be considered in isolation when evaluating the overall performance impact.

\begin{table}[!t]
	\renewcommand{\arraystretch}{1.3}
	\caption{CPU load measurements (in cycles).}
	\label{tab:refcount-t-cpu}
	\footnotesize
	\centering
	\begin{tabular}{c|c|c|c}
			Function		   &  SkyLake	& Sandy Bridge	& Ivy Bridge-EP\\
			\hline
			atomic\_inc()	& 15	& 13	& 10	\\
			refcount\_inc() & 31	& 37	& 31	\\
	\end{tabular}
\end{table}

To gauge the overall performance impact of \refcountt we conducted extensive measurements on the
networking subsystem using the Netperf~\cite{netperf} performance measurement suite. We chose
this subsystem because i) it is known to be performance-sensitive; ii) it has a standardized performance
measurement test suite; and iii) we encountered severe resistance on performance grounds when proposing
to convert this subsystem to use \refcountt. The concerns are well-founded due to the heavy use of reference
counters (e.g., when sharing networking sockets and data) under potentially substantial network
loads. The main challenge when evaluating performance impact on the networking subsystem
is that there are no standardized workloads, and no agreed criteria as to what constitutes ``acceptable'' performance overhead.  
Our test setup consisted of physically connected server and client machines running the
mainline \setype{v4.11-rc8} kernel. 
We measured the real-world performance impact of converting all 78 reference counters in the networking subsystem and networking drivers from \setype{atomic\_t} to \refcountt.

\begin{table}[!t]
	\renewcommand{\arraystretch}{1.3}
	\caption{Netperf refcount usage measurements}
	\label{tab:refcount-t-net}
	\centering
	\footnotesize
	\begin{tabular}{c|c|c|c}
		Netperf Test type		   	& base & refcount & change (stddev)		\\
		\hline
		UDP CPU use (\%) 	& 0.53 		& 0.75				& +42.1\% (34\%)	\\
		TCP CPU use (\%) 	& 1.13  		& 1.28 				& +13.3\% (3\%) \\
		TCP throughput (MB/s)	& 9358 		& 9305 				& -0.6\% (0\%)\\
		TCP throughput (tps)		& 14909 		& 14761				& -1.0\% (0\%)\\
	\end{tabular}
\end{table}

As shown in Table~\ref{tab:refcount-t-net}, our Netperf measurements include CPU
utilization for UDP and TCP streams, and TCP throughput in terms of MB/s and transactions per second (tps).
The results indicate that throughput loss is negligible, but the average processing overhead can be
substantial, ranging from $13\%$ for UDP to $42\%$ for TCP. The
processing overhead can in many situations be acceptable; Desktop systems typically only
use networking sporadically, and when they do, the performance is typically limited by the ISP link
speed, not CPU bottlenecks. In contrast, this might not be the case in servers or embedded systems, and
routers in particular, where processing resources may be limited and networking a major contributor
to system load. However, such systems are typically closed systems, i.e., it is not possible to install
additional applications or other untrusted software, so therefore their attack surface is already reduced.
These considerations are the reason for the \setype{CONFIG\_REFCOUNT\_FULL} kernel configuration option, which can be used to disable the protection offered by \refcountt when the performance overhead is too high, whilst still allowing all other systems to benefit from these new protection mechanisms (which are expected to be enabled by default in the future).

\subsubsection*{Out-of-bounds memory access}

MPXK incurs three types of performance overhead (excluding compile-time overheads). First, the
instrumentation naturally increases CPU utilization and kernel size. Although we do not use the
costly MPX bound storage, some memory will nonetheless be used to store static global and
intermediary bounds. Memory use comparisons, however, indicate that the memory overhead is
negligible. When deploying MPXK over the \setype{xfrm} subsystem, the memory overhead for in-memory
kernel code is $110KB$, which is a $0.7\%$ increase in total size. Kernel image size overhead is
increased by $~45KB$ or $0.6\%$. Some CPU overhead is expected due to the
instrumentation and bound handling. To measure this, we conducted micro-benchmarks on \setype{kmalloc} and
\setype{memcpy}, comparing the performance of MPXK and KASAN, with the results shown in
Table~\ref{tab:mpxk-perf}.

\begin{table*}[!t]
	\renewcommand{\arraystretch}{1.3}
	\caption{MPXK and KASAN CPU overhead comparison.}
	\label{tab:mpxk-perf}
  \footnotesize
	\centering
	\begin{tabular}{l|c|cc|cc}

		\multicolumn{1}{c}{} & \multicolumn{1}{c}{baseline} & \multicolumn{2}{c}{KASAN} & \multicolumn{2}{c}{MPXK} \\
		& time (stddev) & $ns$ diff (stddev) & $\%$ diff & $ns$ diff (stdev) & $\%$ diff \\
		\hline

		\textbf{No bound load.} & & & & & \\
		memcpy, 256 B & $45$ ($0.9$)	& $+85$ ($1.0$) 	& $+190\%$	& $+11$ ($1.2$)		& $+30\%$	\\
		memcpy, 65 kB & $2340$ ($4.3$)	& $+2673$ ($58.1$) 	& $+114\%$	& $+405$ ($5.1$)	& $+17\%$	\\

		\textbf{Bound load needed.} & & & & & \\
		memcpy, 256 B & $45$ ($0.8$)	& $+87$ ($0.9$)		& $+195\%$ 	& $+70$ ($1.5$)		& $+155\%$ 	\\
		memcpy, 65 kB & $2332$ ($5.8$)	& $+2833$ ($28.2$)	& $+121\%$ 	& $+475$ ($15.0$)	& $+20\%$	\\

	\end{tabular}
\end{table*}

To better gauge the actual runtime performance impact, we again applied MPXK to \setype{xfrm} and
measured the impact on an IPSec tunnel where \setype{xfrm} manages the IP package transformations.
For measurements we used Netperf, running five-minute tests for a total of 10 iterations. The test
system was running Ubuntu 16.04 LTS with a v4.8 Linux kernel.
As shown in Table~\ref{tab:mpxk-xfrm-perf}, there is a small reduction in TCP throughput, but the impact on CPU performance is negligible. 

\begin{table}[!t]
	\renewcommand{\arraystretch}{1.3}
	\caption{Netperf measurements over an IPSec tunnel with the \setype{xfrm} subsystem protected by MPXK.}
	\label{tab:mpxk-xfrm-perf}
	\centering
	\footnotesize
	\begin{tabular}{l|c|c|c}
		Netperf test & baseline  & MPXK  & change (stddev) \\
		\hline

		UDP CPU use (\%) 	  &       24.97     &       24.97     &       0.00\%  (0.02)\%        \\
		TCP CPU use (\%) 	  &       25.07     &       25.15     &       0.31\%  (0.29)\%        \\
		TCP throughput (MB/s) &       646.69   &       617.95   &       -4.44\% (4.61)\%        \\
		TCP throughput (tps)  &       1586.79  &       1547.85  &       -2.45\% (1.66)\%        \\

	\end{tabular}
\end{table}

The micro-benchmarks indicate that MPXK, as expected, introduces a measurable performance overhead.
However, compared to KASAN, the performance overhead is relatively small. The large scale tests on
\setype{xfrm} and Netperf confirm that similar performance measurements hold for full real-world systems.
It should further be noted that MPXK is specifically designed for modular deployment, thus ensuring that it can be deployed incrementally without affecting performance sensitive modules or subsystems.

\subsection{Deployability}

\subsubsection*{Reference counter overflows}

Deploying a new data type into a widely used system such as the Linux kernel is a significant real-world challenge. From
a usability standpoint, the API should be as simple, focused, and self-documenting as reasonably
possible. The \refcountt API in principle fulfils these requirements by providing a tightly focused
API with only 14 functions. This is a significant improvement over the 100 functions provided by the
\setype{atomic\_t} type previously used for reference counting. 

Of the 233 patches we submitted, 123 are currently accepted, and it is anticipated that the rest will be accepted in the near future.
The \refcountt type has also been taken into use in independent work by other developers\footnote{\url{http://lkml.org/lkml/2017/6/1/762}}. 
This gives a strong indication that the usability goals are met in practice.
Our contributions also include the Coccinelle pattern used for analysis and detection of potential reference counting schemes that can be converted to \refcountt.
This pattern is currently in the process of being contributed to the mainline kernel, at which point it can be used by individual developers and in various places in the Linux kernel testing infrastructure.

\subsubsection*{Out-of-bounds memory access}
\label{sec:evaluation-deploy-mpxk}

As with MPX, our MPXK design considers usability as a primary design objective. 
It is binary compatible, meaning that it can be enabled for individual translation units. 
It is also source-compatible in that it does not require any changes to source code.
In some cases, pointer bounds checking can interfere with valid pointer arithmetic or other atypical pointer use sometimes seen in high-performance implementations.
However, such compatibility issues are usually only present in architecture-specific implementations of higher-level APIs and can thus be accounted for in a centralized manner by annotating incompatible functions to prevent their instrumentation.
Compatibility issues are also usually found during compile time and are thus easily detected in development. 
MPXK is fully integrated into Kbuild, providing predefined compilation flags for easy deployment.
Using the \setype{MPXK\_AUDIT} parameter, MPXK can also be configured to run in permissive mode where violations are only logged, which is useful for development and incremental deployment.

The MPXK code-base is largely self-contained and thus easy to maintain. 
The only exceptions are the wrapper functions that are used to instrument memory manipulating functions.
However, this is not a major concern because the kernel memory managing and string API is quite stable and changes are infrequent.
The instrumentation is all contained in a GCC-plugin and is thus not directly dependent on GCC internals.
MPXK is thus both easily deployable and easy to maintain.

\section{Related work}
\label{sec:related-work}
Research on memory errors, both temporal and spatial, has long roots both in the industry and in
academia. Many solutions have been presented to eliminate, mitigate and detect memory errors of
various types, but they typically exhibit characteristics that prevent their wide deployment.
Purify~\cite{hastings1991purify}, Shadow Guarding~\cite{patil1995efficient, patil1997low},
SoftBound~\cite{nagarakatte2009softbound}, as well as approaches in~\cite{jones1997backwards},
~\cite{yong2003protecting, xu2004efficient, nethercote2004bounds}, and ~\cite{dhurjati2006backwards},
have non-acceptable run-time performance. CCured~\cite{necula2002ccured} and
Cyclone~\cite{grossman2005cyclone} required source code changes. In addition
CCured~\cite{necula2002ccured}, Cyclone~\cite{grossman2005cyclone} and
approach~\cite{austin1994efficient} are not backward compatible. Moreover, these solutions have
typically focused on user-space and have not been intended or even supported in kernel-space. Some
recent notable exceptions are kCFI~\cite{Rigo} and KENALI~\cite{kenali}, but they have not from an
implementation standpoint targeted actual mainline kernel adoption. To our knowledge, none
of these systems are in production use as run-time security mechanisms. A notable exception is the
\paxgr patches that have pioneered many in-use security mechanisms such as Address Space Layout
Randomization~\cite{aslr2003pax}. \paxgr also includes a
\setype{PAX\_REFCOUNT}~\cite{grsecOnRefcount} feature that prevents reference counter overflows, but
this requires extensive modification of the underlying atomic types. These extensive changes, and a
potential race condition, made this feature unsuitable for mainline kernel
adoption~\cite{grsecOnRefcount}.

Some tools have reached high prominence and active use in the debugging of memory errors.
These include Valgrind~\cite{nethercote2007valgrind} which offers a suite of six tools for debugging,
profiling, and error detection, including the detection of memory errors.
AddressSanitizer~\cite{serebryany2012addresssanitizer} provides something of a gold standard in
run-time memory error detection, but is unsuitable for
run-time production use due to performance overhead. KASAN~\cite{kasan} is
integrated into the mainline Linux kernel and offers similar protections for the kernel, but again
incurs performance overheads unsuitable for most production use cases. 

From a production perspective, much work has focused on preventing the exploitation of memory
errors. Exploitability of buffer overflows, whether stack-based or heap-based, can be limited by
preventing an attacker from misusing overflown data. One early mitigation technique is the
non-executable (NX) bit for stack, heap and data pages~\cite{designer1997linux, team2003pax}. However, this
can be circumvented by using overflows for execution redirection into other legitimate
code memory, such as the C library in so called return-to-libc attacks, or more generally to any
executable process memory in return oriented programming attacks~\cite{krahmer2005x86}. Another
mitigation technique is the use stack canaries that allow the detection of overflows, e.g.,
StackGuard~\cite{cowan1997stackguard} and StackShield~\cite{shield2011stack}. Such detection
techniques can typically be circumvented or avoided using more selective overflows that avoid the
canaries, or by exploiting other callback pointers such as exception handlers. Probabilistic
mitigation techniques such as memory randomization~\cite{aslr2003pax, xu2003transparent} are commonly
used, but have proven difficult to secure against indirect and direct memory leaks that divulge
randomization patterns, not to mention later techniques such as heap
spraying~\cite{ratanaworabhan2009nozzle}.

In contrast to the development/debugging temporal safety tools and the
run-time-friendly mitigation measures, MPXK is a run-time efficient system focused
on the prevention of the underlying memory errors. MPXK is
conceptually similar to previous solutions such as~\cite{kwon2013low, kuvaiskii2017sgxbounds,
necula2002ccured, grossman2005cyclone, nagarakatte2009softbound}. Like these systems, MPXK does not
use fat-pointers, which alter the implementation of pointers to store both the pointer and the
bounds together, but instead preserves the original memory layout. Unlike purely
software-based systems such as SoftBound~\cite{nagarakatte2009softbound}, MPXK has the advantage of
hardware registers for propagating bounds and hardware instructions for enforcing bounds.
HardBound~\cite{devietti2008hardbound} employs hardware support similar to MPXK but has a worst
case memory overhead of almost $200\%$ and has only been simulated on the micro-operation level and
to date lacks any existing hardware support. Unlike MPXK, none of these previous schemes have been designed for use in the kernel.

\section{Discussion}
\label{sec:discussion}
The solutions proposed in this paper and related Linux kernel mainline patches are a step towards better memory
safety in the mainline Linux kernel. One sobering fact when working with production systems and
distributed development communities, such as the Linux kernel, is that there is always
a trade-off between security and deployability. Security solutions outside the mainline kernel, such
as the long-lived PaX/Grsecurity~\cite{grsecurity} patches can thrive, but in order to provide wide-spread security to
a diverse space of devices, it is imperative to achieve integration into the mainline kernel. 

\subsection*{Future work:}
\label{sec:future-work}

Our efforts to convert the remaining reference counters to \refcountt continue, but the initial work has already sparked other implementation efforts and a renewed focus on the problem.
There are active efforts to migrate to the mainline kernel solutions that provide high-performance architecture-independent implementations%
\ifnotanonymous
~\cite{cookSec413}.
\else
~\cite{anonymized}.
\fi

Several MPXK improvements are also left for future work. Our support for bound loading can be
extended with support for other allocators and memory region-based bounds for pointers not
dynamically allocated. More invasive instrumentation could also be used to improve corner cases
where function calls require bound loading. The wrapper implementations could similarly be improved
by more invasive instrumentation, namely by forgoing wrappers altogether and instead employing
direct instrumentation at call sites. This approach is not feasible on vanilla MPX due complications
caused by the bound storage, but would be quite reasonable for MPXK.

\subsection*{MPX: suggestions for improvements:}
\label{sec:mpx-improvements}

While working with \intel MPX and adapting it for kernel use, we have identified some aspects that
limit its usability for this use in particular, but also in more general use cases. One particular
challenge was lacking documentation of both hardware and compiler behavior in atypical cases. In
many cases, such behavior is impossible to determine without either manual run-time testing or
extensive compiler source code analysis. While these limitations might not affect trivial user-space
applications, they are glaring problems when working with performance-critical low-level systems such
as the Linux kernel. This is not ideal for a technology that aims for wide adoption and can
easily become an insurmountable barrier. From a pure hardware perspective, MPX currently provides
only four registers for storing bounds. By increasing the register count, the instrumentation
would be more efficient overall, and in particular function argument bound propagation would benefit by being
able to omit costly bound loads.

\subsection*{Working with Linux kernel community:}
\label{sec:upstream-work}

The path to successful non-trivial mainline kernel patches is often a long and rocky road, especially for
security-related features that have performance implications. Based on our experience in working with 
the Linux kernel developers in this project, we offer the following guidance for other researchers who also
aim to have their solutions merged into the mainline Linux kernel.

\noindent\textbf{Understanding context}. It is not enough to simply read the Linux kernel contribution
guidelines~\cite{kernelPatchDocs} and

follow them when developing your proposal.  It is much more useful to understand the \emph{context}
of the subsystem to which you are attempting to contribute.  By context, we mean the recent history of
the subsystem and planned developments, standard ways in which it is verified and tested, and the
overall direction set by its maintainers. Presenting your contribution by embedding it in this
context will increase the chances of its getting a fair hearing. For example, in our MPXK work,
instead of contributing changes directly to the GCC core, we implemented our GCC instrumentation as
a standalone GCC plugin (Section~\ref{sec:sol-mpxk}) using the GCC kernel plugin framework that
was just recently added to the mainline Linux kernel.  This allowed us to be aligned with the
overall direction of the GCC compiler development. 

\noindent\textbf{Fine-grained configurability}. The Linux kernel runs on very different types of devices and
environments with different threat models and security requirements. Thus any security solution (and
especially those that have performance or usability implications) must not be monolithic but
support the ability to be configured to several different \emph{grades}. For example, we designed
MPXK so that its level of protection can be independently enabled on each compilation unit or
subsystem to protect places where it is most needed (see
Section~\ref{sec:evaluation-deploy-mpxk}). Similarly, in our reference counter protection work, in
addition to having a configuration flag for turning off the protection behind the \refcountt
interface, there is ongoing work to provide an architecture-specific fast assembly
implementation with only slightly relaxed security guarantees%
\ifnotanonymous
~\cite{cookSec413}.
\else
~\cite{anonymized}.
\fi
Also, the conversion to the new
\refcountt API can happen independently for each reference counter. Kernel maintainers can thus
gradually change their code to use \refcountt rather than requiring all kernel code to be changed at
once (in fact, the latter is a major reason for kernel developers turning down the PaX/Grsecurity solution
for reference counters.). 

\noindent\textbf{Timing}. Developing and deploying any new feature that affects more than a single kernel
subsystem takes considerable time. This is due to the number of different people involved in
maintaining various kernel subsystems and the absence of strict organization of the development
process. Researchers should plan for this to make sure enough time is allocated. Also, one has to
take into account various stages of kernel release process to understand when it is the best time to
send your patches for review and get feedback from maintainers. For example, it took us a full year
to reach the current state of reference counter protection work and have 123 out of 233 patches
merged.

Finally, even if a proposed feature is not accepted in the end, both maintainers and security
researchers can learn from the process, which can eventually lead to better
security in the mainline kernel.

\section{Conclusion}
Securing the mainline Linux kernel is a vast and challenging task.
In this paper we present a set of solutions
that on the one hand limit the exploitability of memory errors by eliminating reference counter
overflows and on the other provide hardware assisted solution for enforcing pointer bounds. Both
solutions are aimed at practical deployability, with reference counter protection in particular
already being widely deployed in the mainline kernel. While our solutions arguably exhibit some
limitations, they nonetheless strike a pragmatic balance between deployability and security, thus
ensuring they are in a position to benefit the millions of devices based on the mainline kernel.

\ifnotanonymous
\section*{Acknowledgments}
The authors would like to thank the many mainline Linux kernel developers and maintainers who have
participated in discussions and patiently provided enlightening feedback and suggestions. In short,
for providing us an opportunity to participate and learn about mainline Linux kernel development.

\fi

\bibliographystyle{ACM-Reference-Format}
\bibliography{KernelMemorySafety}

\section*{Appendix}
\begin{lstlisting}[label=cocci, caption=Coccinelle pattern for finding reference counters in the Linux kernel]

// Check if refcount_t type and API should be used
// instead of atomic_t type when dealing with refcounters
// Confidence: Moderate
// URL: http://coccinelle.lip6.fr/
// Options: --include-headers
virtual report
@r1 exists@
identifier a, x;
position p1, p2;
identifier fname =~ ".*free.*";
identifier fname2 =~ ".*destroy.*";
identifier fname3 =~ ".*del.*";
identifier fname4 =~ ".*queue_work.*";
identifier fname5 =~ ".*schedule_work.*";
identifier fname6 =~ ".*call_rcu.*";
@@
(
 atomic_dec_and_test@p1(&(a)->x)          |
 atomic_dec_and_lock@p1(&(a)->x, ...)     |
 atomic_long_dec_and_lock@p1(&(a)->x, ...)|
 atomic_long_dec_and_test@p1(&(a)->x)     |
 atomic64_dec_and_test@p1(&(a)->x)        |
 local_dec_and_test@p1(&(a)->x)
)
...
(
 fname@p2(a, ...); |
 fname2@p2(...);   |
 fname3@p2(...);   |
 fname4@p2(...);   |
 fname5@p2(...);   |
 fname6@p2(...);
)
@script:python depends on report@
p1 << r1.p1;
p2 << r1.p2;
@@
msg = "atomic_dec_and_test variation
      before object free at line \%s."
coccilib.report.print_report(p1[0], 
                  msg \% (p2[0].line))									
@r4 exists@
identifier a, x, y;
position p1, p2;
identifier fname =~ ".*free.*";
@@
(
 atomic_dec_and_test@p1(&(a)->x)          |
 atomic_dec_and_lock@p1(&(a)->x, ...)     |
 atomic_long_dec_and_lock@p1(&(a)->x, ...)|
 atomic_long_dec_and_test@p1(&(a)->x)     |
 atomic64_dec_and_test@p1(&(a)->x)        |
 local_dec_and_test@p1(&(a)->x)
)
...
y=a
...
fname@p2(y, ...);
@script:python depends on report@
p1 << r4.p1;
p2 << r4.p2;
@@
msg = "atomic_dec_and_test variation
      before object free at line \%s."
coccilib.report.print_report(p1[0], 
                  msg \% (p2[0].line))
@r2 exists@
identifier a, x;
position p1;
@@
(
atomic_add_unless(&(a)->x,-1,1)@p1      |
atomic_long_add_unless(&(a)->x,-1,1)@p1 |
atomic64_add_unless(&(a)->x,-1,1)@p1
)
@script:python depends on report@
p1 << r2.p1;
@
msg = "atomic_add_unless"
coccilib.report.print_report(p1[0], msg)
@r3 exists@
identifier x;
position p1;
@@
(
x = atomic_add_return@p1(-1, ...);      |
x = atomic_long_add_return@p1(-1, ...); |
x = atomic64_add_return@p1(-1, ...);
)
@script:python depends on report@
p1 << r3.p1;
@@
msg = "x = atomic_add_return(-1, ...)"
coccilib.report.print_report(p1[0], msg)
\end{lstlisting}

\end{document}